%% file: 0main.tex
\documentclass{article}

\PassOptionsToPackage{numbers, compress}{natbib}
\usepackage[preprint]{neurips_2026}

\usepackage[utf8]{inputenc} % allow utf-8 input
\usepackage[T1]{fontenc}    % use 8-bit T1 fonts
\usepackage{hyperref}       % hyperlinks
\usepackage{url}            % simple URL typesetting
\usepackage{booktabs}       % professional-quality tables
\usepackage{amsfonts}       % blackboard math symbols
\usepackage{nicefrac}       % compact symbols for 1/2, etc.
\usepackage{microtype}      % microtypography
\usepackage{xcolor}         % colors

\usepackage{subfigure}
\usepackage{multirow}
\usepackage{algorithm,algorithmic}
\usepackage{enumitem}
\usepackage{color,xspace}
\usepackage{soul,ulem}
\usepackage{amsmath}
\usepackage{bbm}
\usepackage{colortbl}
\usepackage{tcolorbox}  % 文本加框
\usepackage{tcolorbox}  % 文本加框
\newcommand{\name}{RPORec\xspace}

\newcommand*\samethanks[1][\value{footnote}]{\footnotemark[#1]}

\title{Reinforced Preference Optimization for Reasoning-Augmented Recommendations}

\author{
Jingtong Gao$^1$\thanks{This work was completed at Kuaishou Technology.} \quad Zeyu Song$^2$ \quad Chi Lu$^2$ \quad Xiaopeng Li$^1$ \\
\textbf{Derong Xu}$^1$ \quad \textbf{Maolin Wang}$^1$ \quad \textbf{Peng Jiang}$^2$ \quad \textbf{Kun Gai}$^2$ \quad \textbf{Qingpeng Cai}$^2$\thanks{Corresponding authors.} \quad \textbf{Xiangyu Zhao}$^1$\samethanks[2]\\
$^1$City University of Hong Kong \quad $^2$Kuaishou Technology\\
\texttt{\{jt.g,xiaopli2-c,derongxu2-c,Morin.wang\}@my.cityu.edu.hk}, \texttt{xianzhao@cityu.edu.hk}\\
\texttt{gai.kun@qq.com}, \texttt{\{songzeyu,luchi,jiangpeng,caiqingpeng\}@kuaishou.com}\\
}

\begin{document}

\maketitle

\begin{abstract}
  Recommender systems are critical for delivering personalized content across digital platforms, and recent advances in Large Language Models (LLMs) offer new opportunities to enhance them with richer world knowledge and explicit reasoning capabilities. With the help of reasoning knowledge, recommendations can better infer users’ underlying intents, adapt to evolving preferences, and leverage semantic relationships for improved accuracy and interpretability. However, existing reasoning-based recommendation methods often fail to fully align the LLM’s reasoning process with recommendation-specific objectives due to structural disruption during integration and difficulties in translating free-form generation into accurate item predictions. In this paper, we introduce \name, a reinforced preference optimization framework that unifies an LLM backbone’s reasoning ability with a dedicated recommendation head (Rechead) for precise item retrieval. \name comprises two stages: (1) Reasoning-Augmented Recommendation Modeling, where high-quality Chain-of-Thought (CoT) reasoning is generated and used as auxiliary knowledge to guide the Rechead in learning recommendation-specific representations; and (2) Advanced Reasoning Refinement and Alignment, in which the trained Rechead produces verifiable rewards to fine-tune the LLM backbone via reinforcement learning, enhancing reasoning quality, structural consistency, and task relevance. Extensive experiments on public benchmarks and large-scale online deployments show that \name consistently outperforms state-of-the-art LLM-based recommendation methods, demonstrating the effectiveness of reasoning-augmented recommendation modeling in real-world systems.
\end{abstract}

% \begin{sloppypar}
\input{1introduction}
% \end{sloppypar}
% \input{2preliminary}
\input{3framework}

\input{4experiments}

\input{5application}

\input{6relatedwork}
\input{7conclusion}

\normalem
\bibliographystyle{ACM-Reference-Format}
\bibliography{bibfile}

\newpage
\appendix
\input{8appendix}

% \newpage
% \input{checklist.tex}

\end{document}

%% file: 1introduction.tex
\section{Introduction}
\label{sec:introduction}

Recommender systems have become indispensable in modern digital platforms, delivering personalized content in domains such as e-commerce, streaming media, and social networking~\cite{lu2015recommender,kumar2014survey}. By accurately modeling user preferences, they enhance user experience, sustain engagement, and generate substantial commercial value~\cite{smith2017two,zhao2019deep}. Recently, Large Language Models (LLMs) have shown remarkable abilities in contextual understanding, complex reasoning, and text generation, enabling the interpretation of user intents and the integration of diverse contextual signals~\cite{zhao2023survey,minaee2024large}. Consequently, recommendation research is gradually transitioning from small deep models toward LLM-based approaches that aim to combine the extensive world knowledge of LLMs with high-precision preference modeling~\cite{zhao2024let,wu2024survey}.

% In current studies~\cite{wang2023can, ferrag2025llm}, reasoning is a defining capability that distinguishes LLMs from conventional neural networks. It facilitates the decomposition of complex problems, multi-hop context integration, and construction of coherent inference chains that go beyond surface pattern matching. When applied to recommendations~\cite{zhang2025towards}, reasoning enables LLM-based models to infer the motivations behind user actions, capture preference evolution, and exploit world knowledge to discover semantic relationships between users and items. By generating explicit reasoning steps such as Chains-of-Thought (CoTs), LLMs can produce recommendations that are not only accurate but also transparent, interpretable, and adaptive to evolving user contexts. Leveraging such reasoning-aware modeling thus holds great potential to transform recommender systems into truly human-aligned decision engines and to improve performance.

In recent studies~\cite{wang2023can, ferrag2025llm}, reasoning has emerged as a defining capability that distinguishes LLMs from conventional neural networks. It enables the decomposition of complex problems, multi-hop context integration, and coherent inference beyond surface-level pattern matching. In recommendation~\cite{zhang2025towards}, such reasoning helps LLM-based models infer user intent, track preference evolution, and leverage world knowledge to uncover semantic relations between users and items. By producing explicit reasoning traces such as Chains-of-Thought (CoTs), LLMs can make recommendations that are not only accurate but also more transparent, interpretable, and adaptive to dynamic user contexts. Reasoning-aware modeling therefore holds strong promise for advancing recommender systems.

% Existing reasoning-based recommendation paradigms can be broadly categorized into two groups. 
% (1) Joint Optimization approaches integrate LLM hidden states with recommendation modules for end-to-end learning~\cite{bao2025heterogeneous,you2025r}, attempting to utilize latent reasoning capacity via task-oriented finetuning. However, direct downstream task-oriented gradient updates on LLM hidden states pretrained for general tasks often distort LLMs' explicit reasoning structures, gradually degrading interpretability and reasoning quality as training progresses. 
% (2) Fine-tuned Generative approaches train LLMs to produce recommendations in natural language~\cite{tan2025reinforced}, and could involving CoT-based inference. While these approaches better preserve and exploit explicit reasoning, aligning free-form textual generation with discrete recommended item IDs presents challenges in accuracy and generalization, especially for training-absent items. Even with Semantic IDs (SIDs)\cite{rajput2023recommender,zhang2025reinforced}, inconsistencies between recommendation-specific tokenization and LLM’s vocabulary lead to persistent semantic gaps. Overall, existing methods either bypass explicit reasoning, compromise it under optimization, or fail to align reasoning logic with the structured prediction objectives of recommendation tasks.

Existing reasoning-based recommendation methods can be broadly divided into two categories.
(1) Joint optimization methods integrate LLM hidden states with recommendation modules for end-to-end training~\cite{bao2025heterogeneous,you2025r}, aiming to exploit latent reasoning ability through task-oriented finetuning. Representative methods such as R$^2$ec~\cite{you2025r} jointly optimize reasoning and prediction through hidden-state coupling. While this design tightly connects reasoning and recommendation signals, directly updating hidden states for downstream recommendation objectives may make explicit reasoning harder to preserve, gradually affecting both interpretability and reasoning quality.
(2) Fine-tuned generative methods train LLMs to generate recommendations in natural language~\cite{tan2025reinforced}, potentially with CoT-based inference. Representative methods such as ReRe~\cite{tan2025reinforced} rely on direct generation for recommendation, while LatentR$^3$~\cite{zhang2025reinforced} further shifts reasoning optimization into latent space without preserving explicit CoTs. Although these approaches better preserve or exploit reasoning, aligning free-form text generation for item retrieval remains difficult and may also lead to performance degradation, especially for training-absent items. Even with Semantic IDs (SIDs)~\cite{rajput2023recommender,zhang2025reinforced}, the mismatch between recommendation-specific tokenization and the LLM vocabulary creates a persistent semantic gap. Overall, existing methods still face trade-offs among explicit reasoning preservation, optimization stability, and alignment with the structured retrieval objectives of recommendation.

To bridge these gaps, we propose \textbf{Reinforced Preference Optimization for Reasoning-Augmented Recommendations (\name)}, which combines LLM CoT reasoning with a specialized recommendation head (Rechead) for precise item prediction and retrieval. Instead of coupling the LLM backbone and Rechead through hidden states, \name uses textual outputs as the interface, preserving reasoning integrity while enabling task-specific recommendation modeling and direct item retrieval, thereby reducing semantic mismatch. However, constructing such a reasoning-augmented framework remains non-trivial due to three key challenges: (1) jointly optimizing the LLM backbone and Rechead toward recommendation objectives; (2) preserving and improving the backbone’s reasoning quality; and (3) filtering noisy reasoning context while retaining useful signals for recommendation.

\name tackles these challenges with three key designs. First, we introduce an \textbf{iterative optimization pipeline}: we first train the Rechead with a frozen LLM backbone, and then refine the backbone using verifiable reward signals from the Rechead via Reinforcement Learning with Verifiable Rewards (RLVR)~\cite{guo2025deepseek}. This two-stage design prevents recommendation-specific gradients from directly distorting LLM hidden representations while supporting both accurate recommendation modeling and high-quality reasoning. Second, we develop a \textbf{reasoning-augmented recommendation modeling} framework with a dedicated reasoning-aware Rechead that effectively integrates high-quality reasoning signals into a retrieval-based architecture for fine-grained next-item prediction. Third, we propose an \textbf{reasoning refinement and alignment} strategy based on carefully designed reward functions to improve both reasoning quality and task alignment of the LLM backbone.

In summary, our contributions are:
\begin{itemize}[leftmargin=*]
    \item We introduce \name, a novel framework that integrates high-quality LLM reasoning text into recommendation, enabling reasoning-augmented decision-making.
    \item We design an iterative optimization pipeline with reasoning-augmented recommendation modeling and advanced reasoning refinement and alignment strategies to improve reasoning quality and task alignment for improved performance.
    \item Extensive experiments on public benchmarks and deployment on a global online platform demonstrate that \name significantly outperforms state-of-the-art recommendation methods.
\end{itemize}

%% file: 3framework.tex
\section{\name Framework}
In this section, we provide an overview of \name and detail its key components. We also provide the preliminary knowledge of GRPO and LLM for Recommendations in Appendix~\ref{sec:preliminary}.

% \subsection{Framework Overview}

\begin{figure}[t]
    \centering
    \includegraphics[width=\linewidth]{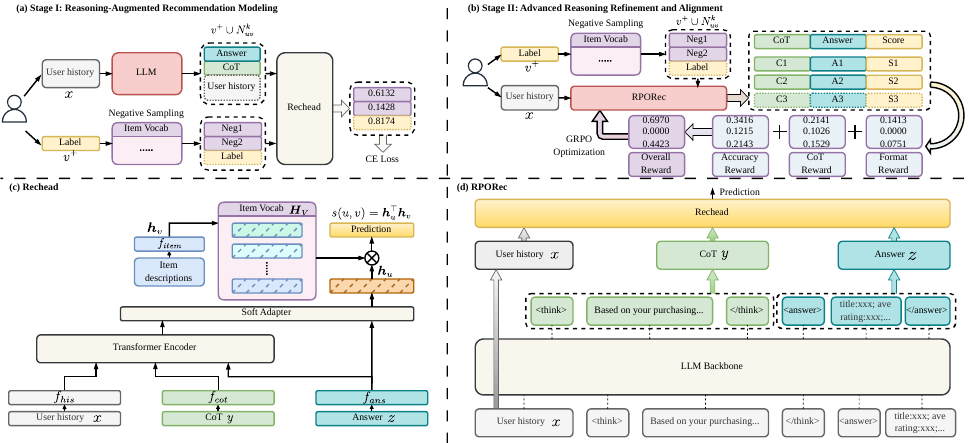}
    \vspace{-7mm}
    \caption{Overview of \name. The full structure of \name is depicted in (d) with detailed Rechead structure in (c). The optimization processes of stage I and II are illustrated in (a) and (b).}
    \vspace{-5mm}
    \label{fig:overview}
\end{figure}

% The proposed framework, \name in Figure~\ref{fig:overview}(d), integrates a reasoning-aware LLM backbone and a reasoning-augmented Rechead (Figure~\ref{fig:overview}(c)) to enhance recommendation quality through controllable reasoning and task alignment. The design follows an iterative optimization principle, alternating between a supervised \textbf{Reasoning-Augmented Recommendation Modeling} stage (Figure~\ref{fig:overview}(a)) for task-specific \textbf{Recommendation head (Rechead)} modeling and an \textbf{Advanced Reasoning Refinement and Alignment} stage (Figure~\ref{fig:overview}(b)) for further refining the \textbf{LLM backbone}'s reasoning ability and aligning the backbone with specific recommendation tasks, ensuring both the reasoning coherence of the LLM backbone and the precision of downstream predictions.
As shown in Figure~\ref{fig:overview}(d), \name consists of a reasoning-aware LLM backbone and a reasoning-augmented Rechead (Figure~\ref{fig:overview}(c)) to improve recommendations through explicit text-based reasoning and direct task-specific item retrieval. The framework follows an iterative optimization paradigm with two stages: Reasoning-Augmented Recommendation Modeling (Figure~\ref{fig:overview}(a)), which learns the recommendation head (Rechead), and Advanced Reasoning Refinement and Alignment (Figure~\ref{fig:overview}(b)), which further improves and aligns the LLM backbone for recommendation tasks. This design jointly preserves reasoning coherence and prediction precision in downstream recommendation.

Specifically, \name couples two core components:
\begin{itemize}[leftmargin=*]
    \item \textbf{Rechead}: provides reasoning augmented recommendation through modeling high-quality LLM reasoning CoTs as auxiliary information for recommendation task-specific modeling and prediction.
    \item \textbf{LLM Backbone}: generates structured outputs containing a reasoning segment with Chain-of-Thoughts (CoTs) and a predicted answer segment describing recommended items with their titles and attributes as shown in  Figure~\ref{fig:casecot}.
\end{itemize}

% During training, the framework operates in two iterative stages:
% \begin{itemize}[leftmargin=*]
%     \item \textbf{Stage~I -- Reasoning-Augmented Recommendation Modeling.} The LLM backbone is frozen, while the Rechead is trained to conduct recommendation modeling using user histories, CoTs, and answer segments as inputs. It first encodes user history, reasoning CoTs, and the output answers using compact, pre-trained sentence transformers, and then applies a lightweight and specialized framework to better utilize effective knowledge in CoTs for reasoning-augmented predictions.
%     \item \textbf{Stage~II -- Advanced Reasoning Refinement and Alignment.} Once the Rechead converges, its output could serve as a stable reward calibrator for fine-tuning the LLM backbone. In this stage, the Rechead is frozen, and the LLM backbone is fine-tuned using reinforcement learning methods in which multiple reward signals are jointly constructed to ensure structural adherence, reasoning fidelity, output utility, and overall recommendation performance.
% \end{itemize}

% Together, these two stages form the foundation of reasoning-augmented recommendations, achieving improved performance with the help of high-quality reasoning CoTs.
During training, the framework operates in two iterative stages:
\begin{itemize}[leftmargin=*]
    \item \textbf{Stage I -- Reasoning-Augmented Recommendation Modeling.} The LLM backbone is frozen, and the Rechead is trained to model recommendation signals from user histories together with the generated CoTs and answer segments.
    \item \textbf{Stage II -- Advanced Reasoning Refinement and Alignment.} The Rechead is then frozen and used as a stable reward source to refine the LLM backbone via reinforcement learning with multiple reward signals.
\end{itemize}

Together, the two stages establish a reasoning-augmented recommendation framework that improves recommendation quality through high-quality CoT reasoning.

\subsection{Reasoning-Augmented Recommendation Modeling}
Leveraging reasoning knowledge for recommendation is non-trivial: directly optimizing recommendation objectives on LLM hidden states may disrupt intrinsic reasoning, while naively encoding CoTs for generic recommenders introduces substantial noise. We therefore propose a lightweight reasoning-augmented recommendation head (Rechead), which treats CoTs as an auxiliary signal in a user-item representation-based retrieval framework without requiring text- or token-level alignment. As shown in Figure~\ref{fig:overview}(a) and (c), Rechead contains four text encoders, a Transformer encoder block, and a soft adapter layer that controls CoT contribution and suppresses irrelevant reasoning in the latent space. 
% Rechead is trained on precomputed CoT-answer pairs generated by the backbone LLM for each training sample.

\subsubsection{Rechead Structure}
As shown in Figure~\ref{fig:overview}(c), Rechead takes the user history $x$, the CoT $y$, and the final answer $z$ (e.g., predicted item title and attributes). Each text is embedded by an encoding function $f$ composed of pre-trained small sentence transformers and feed-forward networks:
\begin{equation}\label{equ:enc}
\boldsymbol{r}_x = f_{his}(x), \quad
\boldsymbol{r}_y = f_{cot}(y), \quad
\boldsymbol{r}_z = f_{ans}(z).
\end{equation}
where $\boldsymbol{r}_x$, $\boldsymbol{r}_y$, $\boldsymbol{r}_z\in \mathbb{R}^{d}$. Items are also encoded from their textual descriptions by a pre-trained small sentence transformer $f_{item}$ into dense vectors $\boldsymbol{h}_v$, which are stored in the item vocabulary matrix $\boldsymbol{H}_{V}\in \mathbb{R}^{|V| \times d}$.

To handle entangled CoTs and answers, or missing answers, Rechead selects a primary representation according to whether a reliable answer $z$ is available:
\begin{equation}\label{equ:select}
\boldsymbol{r}_{sel} =
\begin{cases}
\boldsymbol{r}_z, & \text{if the answer $z$ is parsed successfully},\\
\boldsymbol{r}_x, & \text{otherwise}.
\end{cases}
\end{equation}
This allows Rechead to fall back to direct history modeling when the explicit answer is malformed or absent.

Although CoTs may contain useful reasoning knowledge, their length and limited controllability introduce substantial noise, making raw CoTs ineffective for recommendation. We therefore model interactions among user history, the CoT, and the primary representation with a lightweight transformer encoder, which filters irrelevant content and produces a reasoning-augmented representation:
\begin{equation}\label{equ:trans}
\boldsymbol{r}_{rea} = \mathrm{TransformerEncoder}\!\left(
\left[\boldsymbol{r}_x;\, \boldsymbol{r}_y;\, \boldsymbol{r}_{sel}\right]
\right).
\end{equation}

A gating network then adaptively regulates the contribution of the reasoning-augmented representation to the final representation $\boldsymbol{h}_u \in \mathbb{R}^{d}$:
\begin{equation}\label{equ:gate}
\gamma_0 = \gamma \cdot \sigma (f_{gate}([\boldsymbol{r}_{sel} \parallel \boldsymbol{r}_{rea}])-0.5), \quad
\boldsymbol{h}_u = \gamma_0 \cdot \boldsymbol{r}_{rea} + \boldsymbol{r}_{sel},
\end{equation}
where $\gamma \in (0,1)$ is a hyperparameter, $\sigma$ is the sigmoid function, and $\parallel$ denotes concatenation. This gate curbs noisy or off-topic CoTs while preserving useful relational cues as auxiliary augmentation.

Finally, the recommendation score is computed by a dot product in the shared embedding space, and the item with the highest score is retrieved:
\begin{equation}\label{equ:score}
s(u,v) = \boldsymbol{h}_u^\top \boldsymbol{h}_v .
\end{equation}

\subsubsection{Rechead Optimization}\label{sec:recopt}
As illustrated in Figure~\ref{fig:overview}(a), we precompute the CoT segment $y$ and answer segment $z$ from the LLM for each training sample $[x, v^{+}]$ to construct the training data for Rechead.
% As illustrated in Figure~\ref{fig:overview}(a), Rechead is trained on offline CoT-answer pairs generated for dataset samples, while the LLM backbone in \name remains the same as that in the baselines to ensure a fair comparison.
During inference, predictions are made over the entire item space $\boldsymbol{H}_V$. Exhaustive scoring is expensive during training, so we adopt negative sampling~\cite{ma2024negative} and compute the Cross-Entropy (CE) loss~\cite{mao2023cross} on $k$ sampled negatives $N_{uv}^k$ together with the positive item $v^{+}$ for each training instance:  
\begin{equation}\label{equ:loss}
\mathcal{L}_{\mathrm{rec}} = - \log 
\frac{\exp\big(s(u, v^{+})\big)}{\sum_{v \in \{v^{+}\} \cup N_{uv}^k} \exp\big(s(u,v)\big)}.
\end{equation}  
This formulation enables efficient recommendation training.  

In summary, Rechead uses CoTs as auxiliary reasoning signals, stabilizes prediction through adaptive primary selection, and aligns user vectors with pre-encoded items via efficient contrastive learning. This yields a parameter-efficient and robust module for scalable retrieval under noisy reasoning.

\subsection{Advanced Reasoning Refinement and Alignment}\label{sec:adv-refine}
With Rechead established for task-specific prediction, we next optimize the LLM backbone. Directly using a pre-trained LLM is insufficient because it is often misaligned with recommendation objectives and tends to produce lengthy, low-quality reasoning chains~\cite{jiang2025misaligning}. We therefore fine-tune the backbone with GRPO, using the frozen Rechead as a verifier that provides recommendation-specific feedback. Three complementary reward functions jointly optimize output format, reasoning quality, and recommendation accuracy.

\subsubsection{Format Reward}

To ensure valid Rechead inputs, we adopt a standard format reward~\cite{guo2025deepseek}. The LLM backbone is prompted to generate ``<think>$y$</think><answer>$z$</answer>'', and the format reward is defined as:
Specifically, $r_{fmt}=1.0$ if the format is correct and $r_{fmt}=0.0$ otherwise.
To discourage extraneous content outside the ``think'' and ``answer'' tags, we further introduce a penalty-based reward $r_{clean} = \max(0, 1 - \frac{L_{out}}{\kappa})$,
where $L_{out}$ denotes the length of text outside the designated tags, and $\kappa{=}100$ is a hyperparameter controlling the penalty strength.

\subsubsection{Accuracy Reward}

To align the LLM with recommendation tasks, we design a ranking-based reward. Given a user $u$ with one positive item $v^{+}$ and $k$ sampled negatives $N_{uv}^k$, we compute an NDCG-based reward over the candidate set $\{v^{+}\}\cup N_{uv}^k$ using the Rechead scores $s(u,v)$ (Eq.~\ref{equ:score}). The reward is defined from the NDCG value of the ground-truth item as $r_{ndcg}=\operatorname{NDCG} @ k\left(\operatorname{rank}\left(v^{+}\right)\right)$,
where $\operatorname{rank}()$ denotes the rank of $v^{+}$ over the candidate set.

\subsubsection{CoT Reward}\label{sec:cotreward}
The CoT reward is a central component of \name, because explicit CoTs serve as the key interface between the LLM backbone and Rechead throughout the framework. Accordingly, beyond format correctness and recommendation accuracy, we explicitly optimize the quality of the reasoning trajectory itself, which also distinguishes \name from similar methods that primarily optimize predictions or latent representations. To encourage concise, recommendation-centric reasoning, we use a frozen LLM backbone as a summarizer that maps the original CoT $y$ to a short rationale $\hat{y}$ with designed compression prompts. After obtaining $\hat{y}$, we evaluate the robustness and quality of the original CoT $y$. Specifically, we measure semantic consistency using a pre-trained sentence transformer $e(\cdot)$ and encourage compression via a length-based reward, with $r_{sim}=\mathbf{1}\{\cos(e(y), e(\hat{y})) > \delta\}$ and $r_{comp} = \mathrm{clip}(\frac{|\hat{y}|}{|y|}, 0, 1)$.
where $r_{sim}=1$ when the cosine similarity between the encoded $\hat{y}$ and $y$ exceeds the threshold $\delta$, and $r_{comp}$ is the clipped compression ratio. The similarity reward $r_{sim}$ encourages the original CoT $y$ to retain its core semantics after summarization, while the compression reward $r_{comp}$ favors concise trajectories that resist further simplification, penalizing redundancy. Together, they steer the LLM backbone toward semantically meaningful, information-dense CoTs that would originally require multiple rounds of summarization and simplification.

To further improve CoT quality from a latent decision-making perspective, we introduce an entropy-based reward. Recent work~\cite{wang2025beyond} suggests that token entropy $E_t$ serves as an indicator of generation uncertainty, and that high-entropy tokens—especially those in the top 20\%—are particularly critical to the semantic quality of the generated trajectory. This suggests that high-quality reasoning should retain a small number of informative, uncertain decision points without becoming globally unstable.

To characterize CoT quality via entropy $E$, for a generated CoT $T$, we define the mean entropy $E_\mu=\frac{1}{|T|}\sum_{t\in T}E_t$ and the top-20\% entropy average $E_{20\%}$. The entropy reward is then defined as $r_{\text{ent}} = E_{20\%} - E_\mu$, rewarding trajectories with salient high-information tokens (i.e., higher $E_{20\%}$) while penalizing excessive overall uncertainty (i.e., higher $E_\mu$).

Finally, we aggregate these signals for GRPO training while requiring a non-zero format reward $r_{fmt}$ to enforce strict format generation:
\begin{equation}\label{equ:ra-total}
r = r_{fmt} \cdot (\alpha_0 r_{fmt} + 
 \alpha_1 r_{clean} + \alpha_2 r_{ndcg} + \alpha_3 r_{sim} + \alpha_4 r_{comp} + \alpha_5 r_{ent}).
\end{equation}

\subsection{Iterative Optimization}
To align the components in \name with the recommendation objective while avoiding unstable rewards from simultaneously optimizing multiple modules with a non-stationary Rechead, we adopt a two-stage iterative strategy. We first freeze the LLM backbone and train Rechead for the downstream recommendation task. We then freeze Rechead and use it to provide stable, task-specific feedback for LLM reasoning refinement and alignment. The overall optimization procedure of \name is summarized in Appendix~\ref{sec:alg}.

%% file: 4experiments.tex
\section{Experiments}\label{sec:exp}

In this section, we conduct experiments on three public datasets to investigate the following questions:
\begin{itemize}[leftmargin=*]
    \item \textbf{RQ1:} How does \name perform in comparison with different recommendation baselines?
    \item \textbf{RQ2:} What's the contribution of the designed modules and structures in making recommendations?
    \item \textbf{RQ3:} Could \name actually amplify LLM's reasoning quality and its effect in recommendation modeling?

\end{itemize}

In the following subsections, we first describe the evaluation settings. Afterward, we would answer the corresponding questions based on a brief review of our experiment results.

\subsection{Experimental Setup}
We evaluate \name on three Amazon datasets: Musical Instruments, CDs and Vinyl, and Video Games, against representative traditional, LLM-based, and RL-based baselines. Following prior work~\cite{tan2025reinforced,you2025r} and industrial experiences, a small-size LLM, Qwen3-0.6B~\cite{qwen3technicalreport}, is applied as the LLM backbone for \name and all LLM-based baselines to meet the efficiency requirements of recommendation tasks. Detailed dataset statistics, preprocessing settings, baseline descriptions, and implementation details are given in Appendix~\ref{sec:detailed-exp}.

\subsection{Overall Performance (RQ1)}

This subsection gives an overall comparison between \name and different baselines. The results are depicted in Table~\ref{tab:baseline}. From this we can conclude that: 

% Table generated by Excel2LaTeX from sheet 'overall'
\begin{table*}[thbp]
  \centering
  \caption{Performance comparison of \name and baselines. Boldface denotes the highest performance, and underline indicates the second-best performance. ``\textbf{{\Large *}}'' indicates the statistically significant improvements (i.e., two-sided t-test with $p<0.05$) over the second-best baseline.}
  \setlength{\tabcolsep}{2pt}     % 默认通常约 6pt，可适当压缩
  \renewcommand{\arraystretch}{1.2}  % 从 1.5 降下来
  \resizebox{\textwidth}{!}{
    \begin{tabular}{c|cccccc|cccccc|cccccc}
    \toprule
    \multirow{2}[4]{*}{Method} & \multicolumn{6}{c|}{Musical Instruments}      & \multicolumn{6}{c|}{CDs and Vinyl}            & \multicolumn{6}{c}{Video Games} \\
\cmidrule{2-19}          & {H@5} & {N@5} & {H@10} & {N@10} & {H@20} & {N@20} & {H@5} & {N@5} & {H@10} & {N@10} & {H@20} & {N@20} & {H@5} & {N@5} & {H@10} & {N@10} & {H@20} & {N@20} \\
    \midrule
    GRU4Rec & 0.0101  & 0.0063  & 0.0187  & 0.0091  & 0.0320  & 0.0125  & 0.0044  & 0.0028  & 0.0061  & 0.0034  & 0.0088  & 0.0041  & 0.0145  & 0.0096  & 0.0237  & 0.0126  & 0.0347  & 0.0153  \\
    Caser & 0.0121  & 0.0076  & 0.0215  & 0.0106  & 0.0368  & 0.0144  & 0.0045  & 0.0029  & 0.0066  & 0.0036  & 0.0099  & 0.0044  & 0.0150  & 0.0100  & 0.0236  & 0.0127  & 0.0360  & 0.0158  \\
    SASRec & 0.0149  & 0.0096  & 0.0252  & 0.0128  & 0.0394  & 0.0163  & 0.0092  & 0.0056  & 0.0145  & 0.0073  & 0.0212  & 0.0093  & 0.0227  & 0.0145  & 0.0364  & 0.0191  & 0.0563  & 0.0244  \\
    TIGER & 0.0158  & 0.0099  & 0.0243  & 0.0123  & 0.0381  & 0.0158  & 0.0071  & 0.0043  & 0.0105  & 0.0052  & 0.0156  & 0.0065  & 0.0165  & 0.0082  & 0.0245  & 0.0114  & 0.0382  & 0.0154  \\
    BIGRec & 0.0132  & 0.0082  & 0.0228  & 0.0117  & 0.0367  & 0.0152  & 0.0058  & 0.0035  & 0.0095  & 0.0047  & 0.0123  & 0.0051  & 0.0183  & 0.0091  & 0.0267  & 0.0124  & 0.0359  & 0.0145  \\
    D$^3$    & 0.0162  & 0.0101  & 0.0248  & 0.0126  & 0.0395  & 0.0164  & 0.0074  & 0.0045  & 0.0105  & 0.0052  & 0.0164  & 0.0068  & 0.0201  & 0.0100  & 0.0272  & 0.0127  & 0.0399  & 0.0161  \\
    S-DPO & 0.0167  & 0.0104  & 0.0239  & 0.0122  & 0.0392  & 0.0163  & 0.0115  & 0.0070  & 0.0198  & 0.0099  & 0.0310  & 0.0129  & 0.0229  & 0.0114  & 0.0334  & 0.0156  & 0.0468  & 0.0189  \\
    SPRec  & \underline{0.0226}  & \underline{0.0141}  & 0.0281  & 0.0143  & 0.0436  & 0.0181  & 0.0130  & 0.0079  & 0.0216  & 0.0108  & 0.0325  & 0.0135    & \textbf{0.0328} & \textbf{0.0168} & \underline{0.0453}  & \underline{0.0211}  & 0.0562  & 0.0227  \\
    R$^2$ec  & 0.0217  & 0.0135  & 0.0306  & 0.0156  & \underline{0.0486}  & \underline{0.0202}  & 0.0114  & 0.0069  & 0.0190  & 0.0095  & 0.0320  & 0.0133  & 0.0296  & 0.0147  & 0.0403  & 0.0188  & 0.0475  & 0.0192  \\
    ReRe  & 0.0215  & 0.0134  & \underline{0.0318}  & \underline{0.0162}  & 0.0484  & 0.0201  & 0.0140  & 0.0085  & \underline{0.0224}  & \underline{0.0115}  & 0.0315  & 0.0131  & 0.0300  & 0.0149  & 0.0442  & 0.0206   & 0.0548  & 0.0235 \\
    LatentR$^3$ & 0.0220  & 0.0137  & 0.0295  & 0.0155  & 0.0465  & 0.0195  & \underline{0.0148}  & \underline{0.0090}  & \underline{0.0224}  & 0.0112  & \underline{0.0354}  & \underline{0.0147}  & 0.0285  & 0.0142  & 0.0418  & 0.0198  & \underline{0.0617}  & \underline{0.0249}  \\
    % RPORec-1 & 0.0157 & 0.0098 & 0.0229 & 0.0117 & 0.0345 & 0.0143 & 0.0127 & 0.0066 & 0.0194 & 0.009 & 0.0305 & 0.012 & 0.0221 & 0.011 & 0.0307 & 0.0143 & 0.0401 & 0.0162 \\
    RPORec & \textbf{0.0248*} & \textbf{0.0155*} & \textbf{0.0348*} & \textbf{0.0178*} & \textbf{0.0531*} & \textbf{0.0220*} & \textbf{0.0190*} & \textbf{0.0101*} & \textbf{0.0288*} & \textbf{0.0131*} & \textbf{0.0517*} & \textbf{0.0185*} & \underline{0.0326}  & \underline{0.0162} & \textbf{0.0478*} & \textbf{0.0223*} & \textbf{0.0655*} & \textbf{0.0275*} \\
    \midrule
    Improve (\%) & 9.73\% & 9.93\% & 9.43\% & 9.88\% & 9.26\% & 8.91\% & 28.4\% & 12.22\% & 28.57\% & 13.91\% & 46.05\% & 25.85\% & - & - & 5.52\% & 5.69\% & 6.16\% & 10.44\% \\
    \bottomrule
    \end{tabular}%
    }
    \vspace{-2mm}
  \label{tab:baseline}%
\end{table*}%

\begin{itemize}[leftmargin=*]
    \item Traditional models (GRU4Rec, Caser, SASRec) provide strong baselines, with SASRec performing best due to self-attention. Yet without explicit reasoning, they struggle to capture nuanced and evolving user preferences.

    \item Joint optimization approaches (TIGER, BIGRec, D$^3$) yield moderate gains, especially on sparse datasets CDs and Vinyl. This suggests that direct token-level optimization can distort reasoning, while TIGER's semantic ID generation fails to fully exploit LLM's pretrained reasoning knowledge.

    \item Fine-tuned generative and RL-based methods (S-DPO, SPRec, R$^2$ec, ReRe, LatentR$^3$) perform better overall, confirming the value of reasoning-aware modeling. However, ReRe and LatentR$^3$ may incur performance loss during aggregation, especially when candidate items are absent from training, while hidden-state coupling in R$^2$ec and latent-only optimization in LatentR$^3$ may make explicit reasoning harder to preserve. This trend is consistent with the design of \name, which preserves explicit CoTs while decoupling them from final retrieval through a specialized recommendation head.

    \item \name outperforms all baselines across datasets and most metrics, validating the combination of explicit CoT reasoning and a specialized recommendation head. By preserving reasoning in text and using iterative optimization with verifiable rewards, \name better aligns LLM reasoning with recommendation objectives and produces stronger, more interpretable recommendations.
    
\end{itemize}

% \begin{figure}[t]
% 	\centering
%         \subfigure[H@10.]{
% 		\label{fig:ab1}
% 		\includegraphics[width=0.8\linewidth]{fig/ablation_study_H@10.pdf}}
%         \vspace{-3mm}
        
%         \subfigure[N@10.]{
% 		\label{fig:ab2}
% 		\includegraphics[width=0.8\linewidth]{fig/ablation_study_N@10.pdf}}
%         \vspace{-3mm}
% \caption{Ablation study.}
% \label{fig:ablation}
% \end{figure}

\subsection{Ablation Study (RQ2)}
To assess the contribution of \name's components, we conduct an ablation study with the following variants on CDs and Vinyl dataset and report H@10 and N@10:

\begin{itemize}[leftmargin=*]
    \item \textbf{\name:} The complete model.
    \item \textbf{-cot:} Removes CoT inputs and reasoning-augmented modeling in the Rechead during Stage I.
    \item \textbf{-I:} Removes the entire Stage I training, and performs item retrieval directly based on the LLM backbone outputs.
    \item \textbf{-fmt / -clean / -sim / -comp / -ent:} Removes the corresponding reward term from Stage II.
    \item \textbf{-II:} Removes the entire Stage II, without fine-tuning the LLM backbone.
\end{itemize}

\begin{figure}[h]
    \centering
    \subfigure[H@10.]{
        \label{fig:ablation}
    \includegraphics[width=0.44\linewidth]{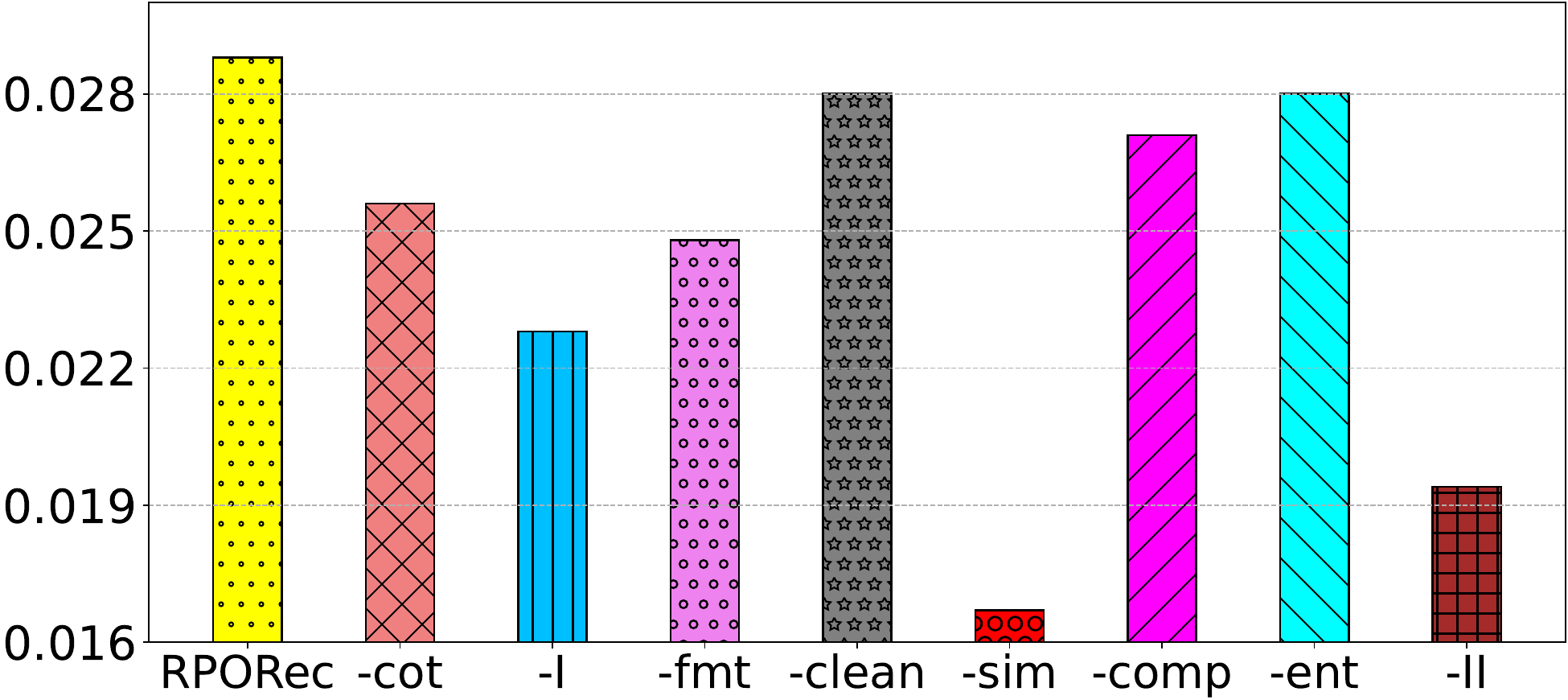}}
    \subfigure[N@10.]{
        \label{fig:ablation1}
    \includegraphics[width=0.44\linewidth]{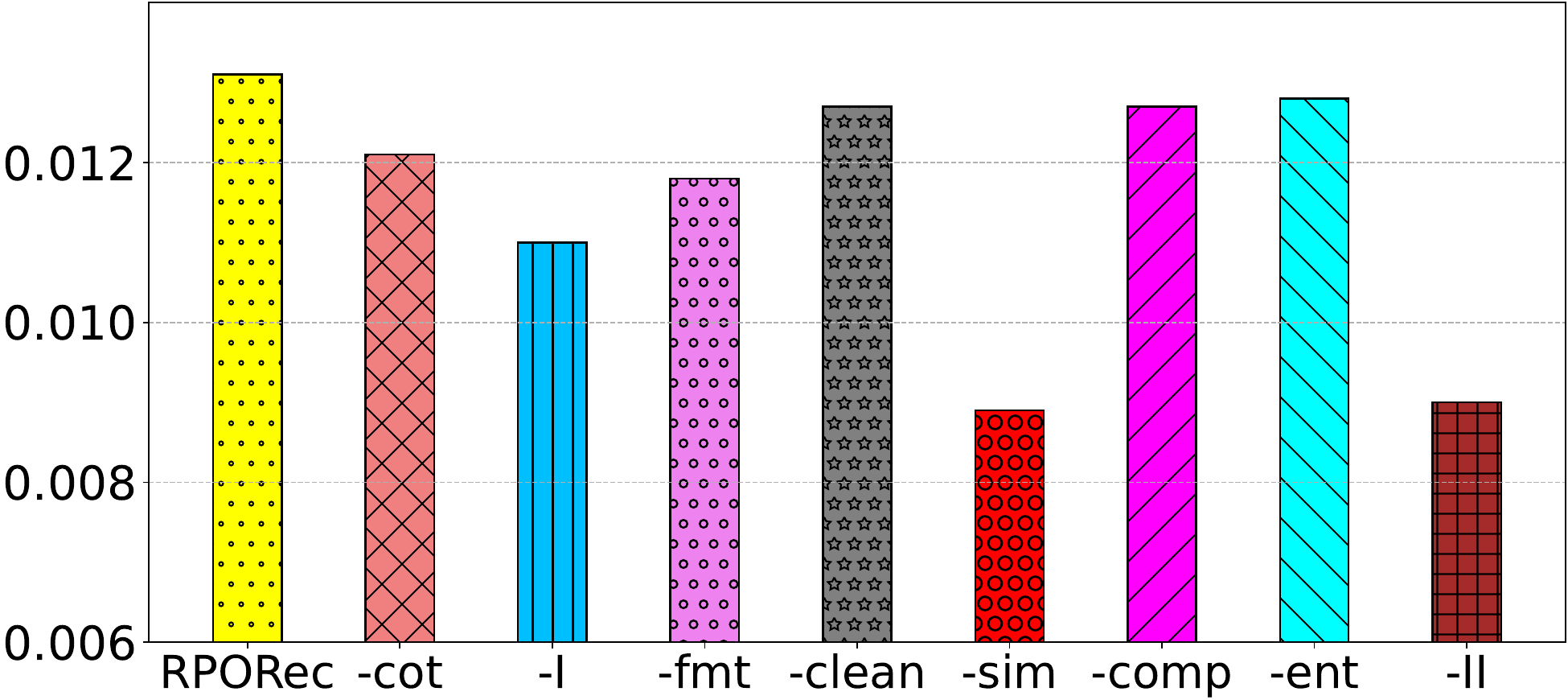}}
    \vspace{-5mm}
    \caption{Ablation study.}
    \vspace{-3mm}
    \label{fig:abl}
\end{figure}

% Based on the results in Figure~\ref{fig:ablation}, we observe that Stage-II optimization is the most critical component, as removing it causes the largest performance collapse, demonstrating that reasoning alignment is indispensable for adapting LLM reasoning to recommendation tasks. The Rechead also plays a vital role by bridging LLM reasoning with discrete item prediction through its reasoning-augmented architecture and adaptive gating mechanism. The modeling of CoT in Rechead contributes meaningfully by providing high-quality, valuable auxiliary signals beyond surface patterns. Other components including format reward (-FR) and CoT reward (-CR) show moderate but consistent improvements, validating their effectiveness in enhancing reasoning quality.
Based on the ablation results in Figure~\ref{fig:ablation} and~\ref{fig:ablation1}, we could draw the following conclusions:

% \begin{figure}[t]
% 	\centering
% \includegraphics[width=0.7\linewidth]{fig/ablation_H10.pdf}
% \vspace{-4mm}
% \caption{Ablation study H@10.}
% \vspace{-6mm}
% \label{fig:ablation}
% \end{figure}

% \begin{figure}[t]
% 	\centering
% \includegraphics[width=0.7\linewidth]{fig/ablation_N10.pdf}
% \vspace{-4mm}
% \caption{Ablation study N@10.}
% \vspace{-3mm}
% \label{fig:ablation1}
% \end{figure}

\begin{itemize}[leftmargin=*]
    \item \textbf{Stage I components.} Removing CoT-aware modeling (\textbf{-cot}) causes a clear performance drop, showing that Stage I benefits substantially from explicit CoT utilization in Rechead rather than relying only on the backbone outputs.

    \item \textbf{Role of Rechead.} Removing Stage I and Rechead (\textbf{-I}) markedly degrades performance, confirming that free-form backbone outputs alone are insufficient for accurate retrieval and that Rechead is essential for bridging reasoning and recommendation.

    \item \textbf{Stage II rewards.} Removing any reward term in Stage II lowers performance, showing their complementarity. In particular, \textbf{-sim} causes the largest drop, indicating that similarity reward is crucial for keeping CoT optimization semantically grounded; without it, $r_{comp}$ reward may still shorten reasoning, but is more likely to drive it away from valid recommendation-relevant content.

    \item \textbf{Role of Stage II.} Removing the entire Stage II (\textbf{-II}) sharply reduces performance, demonstrating that RL-based refinement is indispensable for recommendation alignment.
\end{itemize}

Moreover, we further evaluate \name with Llama3.2-1B~\cite{dubey2024llama} on CDs and Vinyl (Table~\ref{tab:backbone}) to test \name's generalization across different LLM backbones. Results demonstrate consistent performance across the two base models, confirming \name's robustness and generalizability. 

\begin{table}[h]
  \centering
  \vspace{-2mm}
  \caption{Backbone generalization analysis.}
  \vspace{-1mm}
    \begin{tabular}{c|cc}
    \toprule
    Model & H@10 & N@10 \\
    \midrule
    \name (Qwen3-0.6B) & 0.0288  & 0.0131  \\
    \name (Llama3.2-1B) & 0.0294 & 0.0128 \\
    \bottomrule
    \end{tabular}%
    \vspace{-5mm}
  \label{tab:backbone}%
\end{table}%

% \begin{figure*}[t]
% 	\centering
% \includegraphics[width=\linewidth]{fig/case-cot.png}
% \vspace{-7mm}
% \caption{Case study on reasoning CoT quality. finding better examples with <answer> segment}
% \vspace{-3mm}
% \label{fig:casecot}
% \end{figure*}

% \begin{figure}[t]
% 	\centering
% \includegraphics[width=0.8\linewidth]{fig/CoT_reward_comparison.pdf}
% \vspace{-3mm}
% \caption{CoT length comparison.}
% \vspace{-6mm}
% \label{fig:tokenlen}
% \end{figure}

\subsection{Case Study (RQ3)}
\begin{figure}[h]
	\centering
\includegraphics[width=0.95\linewidth]{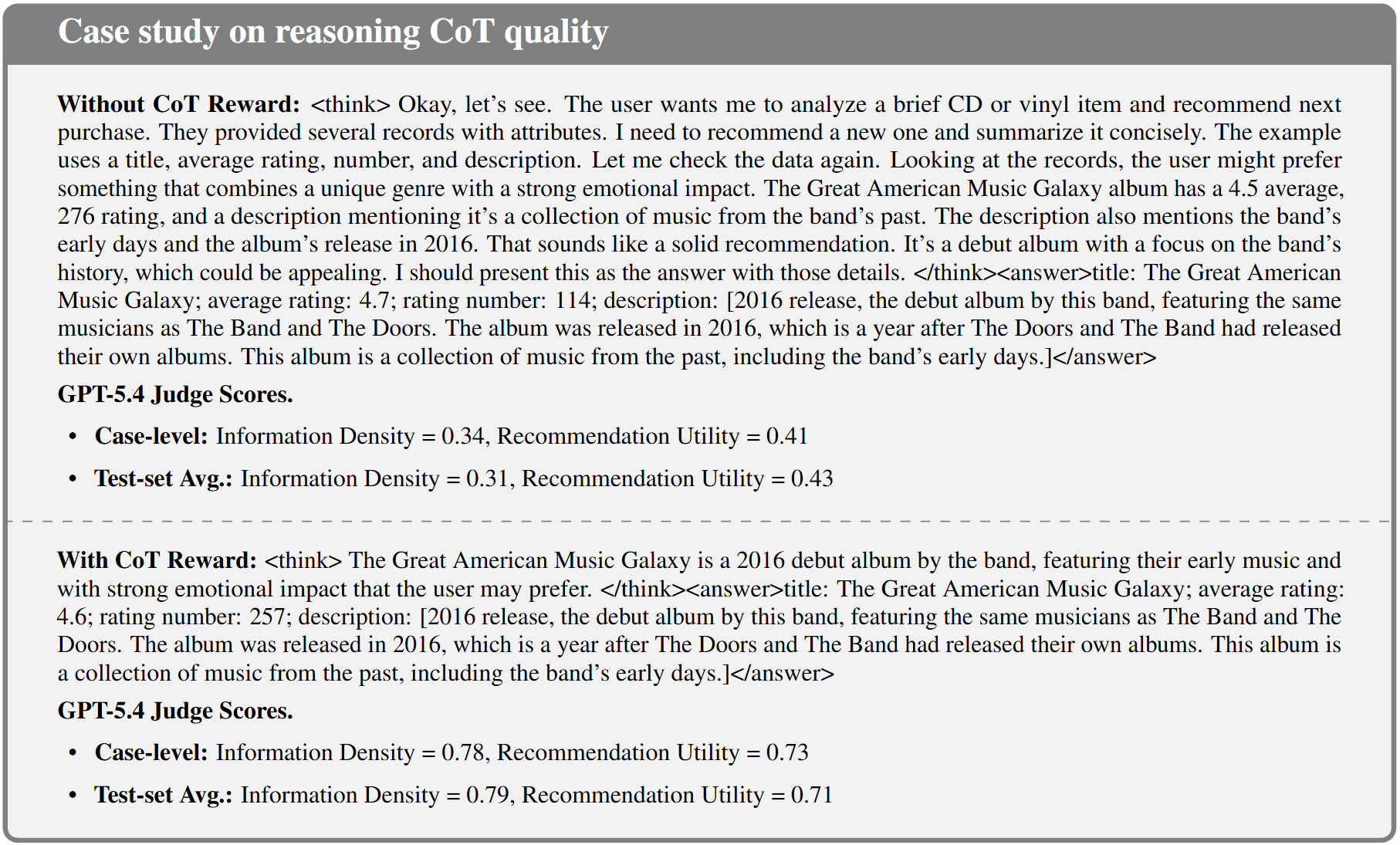}
\vspace{-3mm}
\caption{Case study on reasoning CoT quality.}
\vspace{-4mm}
\label{fig:casecot}
\end{figure}

\begin{figure}[h]
	\centering
    \vspace{-2mm}
\includegraphics[width=0.55\linewidth]{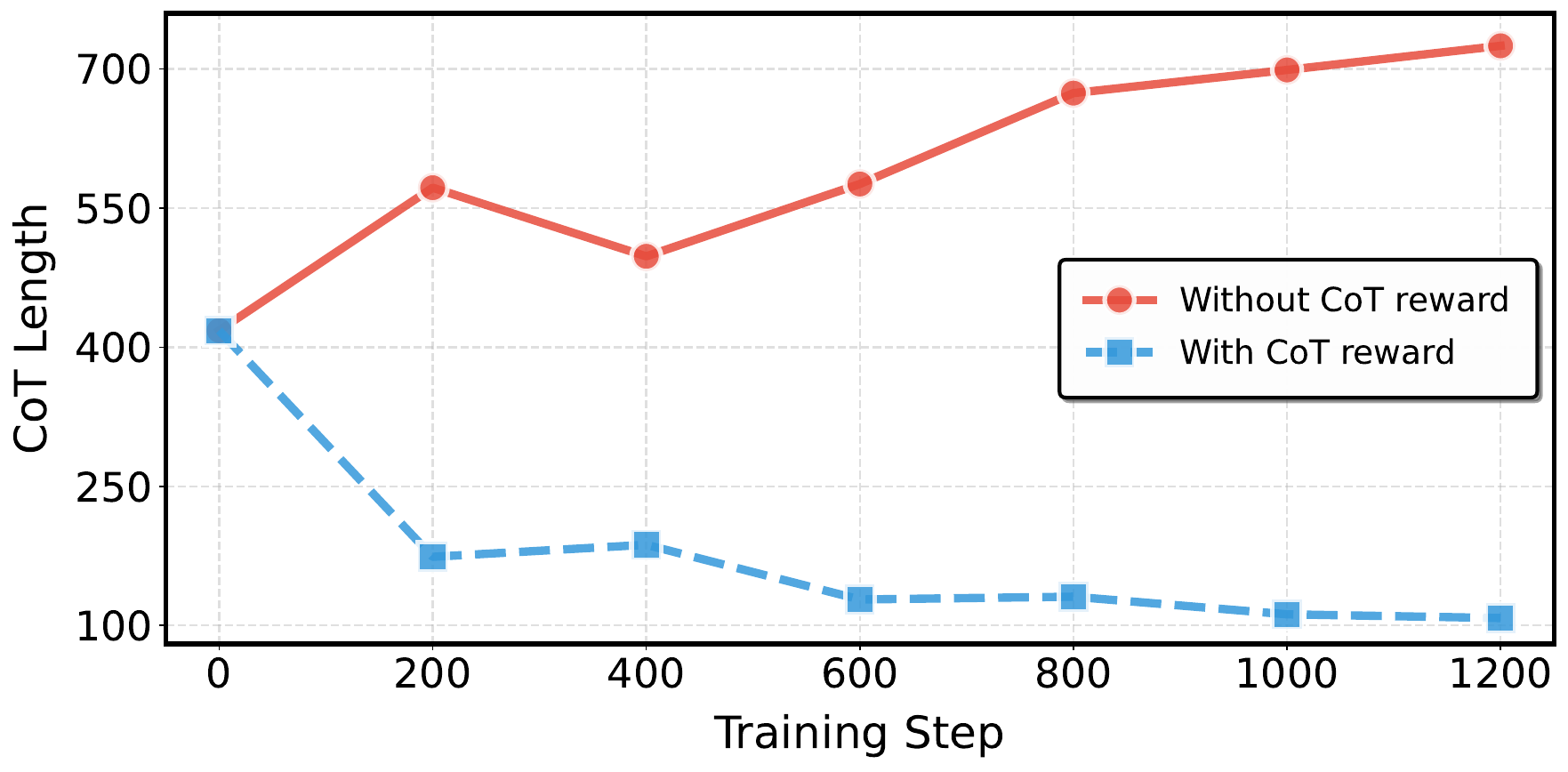}
\vspace{-3mm}
\caption{CoT length comparison.}
\vspace{-6mm}
\label{fig:tokenlen}
\end{figure}

To provide an intuitive analysis of \name’s effect on explicit reasoning, we conduct a case study on the CDs and Vinyl dataset. As the main reasoning enhancement in \name comes from the CoT Reward in Section~\ref{sec:cotreward}, we compare the LLM backbone’s reasoning content and average output length before and after applying this reward. Representative examples are shown in Figure~\ref{fig:casecot}. To further assess CoT quality, we use GPT-5.4~\cite{GPT54_2026} as an LLM judge and evaluate generated CoTs from two perspectives: \textbf{Information Density}, which measures whether the reasoning is concise yet informative rather than redundant or noisy, and \textbf{Recommendation Utility}, which measures whether the reasoning provides evidence that is useful for identifying the correct recommended item; both are scored from 0 to 1.
We also provide the averaged output length comparison in Figure~\ref{fig:tokenlen}.

% \begin{figure}[t]
%     \centering
%     \includegraphics[width=0.87\linewidth]{fig/case study.png}
%     \vspace{-4mm}
%     \caption{Case study on reasoning CoT quality.}
%     \vspace{-4mm}
%     \label{fig:casecot}
% \end{figure}

From Figure~\ref{fig:casecot}, we observe that without CoT reward, the model tends to produce verbose reasoning that repeats item metadata (e.g., release year and background description) and introduces weak speculative statements about user preference, which dilutes recommendation-relevant evidence. After applying CoT reward, the reasoning becomes markedly shorter and more focused, retaining only the core cues that support the recommendation decision. This qualitative change is also reflected by the GPT-5.4 judge scores: for the illustrated case, \textbf{Information Density} improves from 0.34 to 0.78 and \textbf{Recommendation Utility} improves from 0.41 to 0.73; similar gains are also observed on the test-set averages (0.31 $\rightarrow$ 0.79 and 0.43 $\rightarrow$ 0.71, respectively). Concurrently, as shown in Figure~\ref{fig:tokenlen}, the inference length decreases effectively with training, substantially reducing noise in the reasoning trajectory. This reduction not only improves inference quality but also enhances generation efficiency.

%% file: 5application.tex
\section{Online Application}

To validate the effectiveness of \name in real-world deployment, we integrated it into a large-scale industrial advertising system and conducted rigorous online A/B testing.

\begin{figure}[t]
    \centering
    \includegraphics[width=0.8\linewidth]{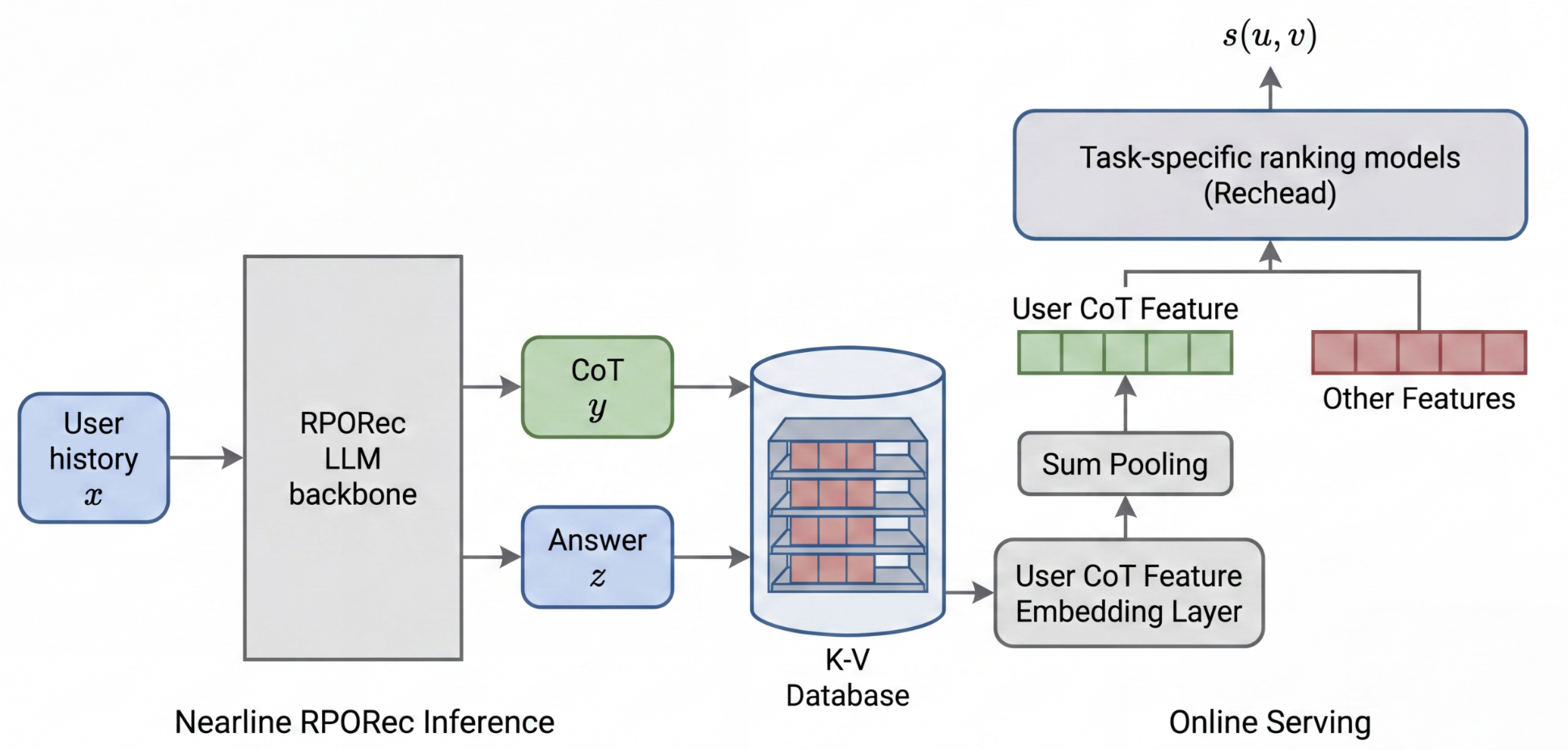}
    \vspace{-3mm}
    \caption{Online deployment architecture of \name.}
    \vspace{-5mm}
    \label{fig:online_arch}
\end{figure}

Figure~\ref{fig:online_arch} illustrates the online deployment architecture. Given the stringent constraints on computational cost, inference latency, and feature interaction complexity in production environments, we deploy the LLM backbone as a nearline user understanding module, while using task-specific online ranking models as Recheads for fine-grained modeling. Specifically, the \name LLM backbone analyzes user profile attributes and historical behavior to predict user interests. We then extract the CoT and answer segments from the generated responses and store them in a key-value (K-V) database. During online serving, the CoT tokens are embedded into dense vectors and incorporated as auxiliary user features into downstream ranking models (i.e., Recheads), together with other online features.

The A/B test ran for 7 days with 10\% traffic allocation, covering approximately 40 million users and 2.1 billion ad impressions. The baseline is a state-of-the-art ranking model currently deployed in production, incorporating extensive handcrafted features and GSU-ESU modules~\cite{pi2020search}, with 0.8 billion sparse and 0.2 billion dense parameters. Online applications demonstrate that integrating \name achieved a \textbf{1.348\%} lift in Revenue and a \textbf{1.058\%} increase in Advertiser Value (ADVV)~\cite{chai2025longer}.

%% file: 6relatedwork.tex
\section{Related Works}

Recommender systems have evolved from collaborative filtering to deep learning~\cite{hidasi2015session,tang2018personalized}, and recently entered a new paradigm leveraging LLMs' world knowledge. TIGER~\cite{rajput2023recommender} pioneers this shift through generative retrieval that autoregressively decodes Semantic IDs (SIDs) from collaborative signals, enabling LLMs to directly generate recommendations. Subsequent work enhances generation quality through debiasing~\cite{bao2024decoding} and preference alignment via DPO variants~\cite{chen2024softmax,gao2025sprec} and RLVR-based optimization~\cite{tan2025reinforced}. With RLVR, LLM-based methods have progressed toward reasoning-aware recommendations. R$^2$ec~\cite{you2025r} employs a dual-head architecture jointly generating reasoning and predictions via hidden-state updates. LatentR$^3$~\cite{zhang2025reinforced} optimizes compact latent reasoning in hidden space using modified GRPO without explicit text generation. ReRe~\cite{tan2025reinforced} directly conducts recommendations through constraint decoding. However, applying reasoning-augmented LLMs poses format and task alignment challenges. Hidden-state methods suffer coarse-grained updates that inadequately refine explicit reasoning, while generative approaches face semantic gaps between LLM outputs (text or pretrained tokens) and recommendation objectives (discrete item IDs or SIDs). \name addresses these limitations through a decoupled two-stage framework with an LLM backbone for reasoning and a specialized Rechead for recommendations. This design enables RLVR to align explicit reasoning CoT at the text level, enhancing reasoning quality while effectively leveraging CoT for recommendation modeling and item retrieval in Rechead without semantic gap interference.

%% file: 7conclusion.tex
\section{Conclusion}

We introduce \name, a reinforcement learning framework that systematically addresses the challenge of integrating high-quality LLM reasoning into recommender systems through three methodological advances. First, our iterative optimization pipeline decouples reasoning generation from recommendation prediction, enabling joint refinement while preserving reasoning integrity and ensuring task-specific alignment. Second, our reasoning-augmented Rechead effectively exploits CoT knowledge through lightweight transformers and adaptive gating mechanisms, filtering noisy content while distilling informative signals for enhanced recommendation precision. Third, we propose advanced reasoning refinement and alignment through RLVR-based optimization, incorporating format adherence, accuracy signals, and CoT quality improvements to produce concise, task-relevant reasoning trajectories for improved performance. Extensive experiments demonstrate that \name significantly outperforms state-of-the-art baselines across public benchmarks, while deployment in a large-scale industrial system achieved 1.348\% Revenue lift and 1.058\% Advertiser Value increase, establishing its effectiveness in delivering superior performance, interpretability, and practical applicability for reasoning-augmented recommendations.

%% file: 8appendix.tex
\input{2preliminary}

\begin{algorithm}[ht]
    \caption{\label{alg:overall} Iterative optimization algorithm of \name}
    \raggedright
    {\bf Input}: A training dataset $\mathcal{D}=\{(x_j, v_j^+)\}_{j=1}^{|\mathcal{D}|}$, 
    where $x_j$ is the user history and $v_j^+$ is the ground-truth next item. A pre-trained LLM backbone $\boldsymbol{\Theta}_{LLM}$ and an initialized Rechead $\boldsymbol{\Theta}_{R}$.
    {\bf Output}: Optimized \name model with $(\boldsymbol{\Theta}_{LLM}^*, \boldsymbol{\Theta}_{R}^*)$.

    {\noindent \bf \hspace{4.5mm} Stage I -- Reasoning-Augmented Recommendation Modeling:}
    \begin{algorithmic}[1]

        \STATE Generate fixed $y_j$ and $z_j$ for each $(x_j, v_j^+)$ in $\mathcal{D}$ with $\boldsymbol{\Theta}_{LLM}$ once.
        \FOR{epoch $=1$ \textbf{to} max\_epochs}
            \FOR{each mini-batch $B \subset \mathcal{D}$}
                \STATE Forward the $(x_j, y_j, z_j) \in B$ through Rechead via Equation~\eqref{equ:enc}, ~\eqref{equ:select}, ~\eqref{equ:trans}, ~\eqref{equ:gate}, ~\eqref{equ:score}.
                \STATE Sample $k$ negative items $N_{uv}^k$.
                \STATE Update $\boldsymbol{\Theta}_{R}$ via Equation~\eqref{equ:loss}.
            \ENDFOR
            \IF{Rechead converged}
                \STATE Obtain and freeze $\boldsymbol{\Theta}_{R}^*$, and proceed to Stage II.
                \STATE \textbf{break}
            \ENDIF
        \ENDFOR
        \newline{\bf Stage II: Advanced Reasoning Refinement and Alignment:}
        \FOR{epoch $=1$ \textbf{to} max\_epochs\_RL}
            \FOR{each mini-batch $B \subset \mathcal{D}$}
                \STATE Generate $y_j$ and $z_j$ for each $(x_j, v_j^+)$ with $\boldsymbol{\Theta}_{LLM}$.
                \STATE Forward $(x_j, y_j, z_j) \in B$ through Rechead to obtain $s(u,v)$ in Equation~\eqref{equ:score}.
                \STATE Compute $r_{fmt}$, $r_{clean}$, $r_{ndcg}$, $r_{sim}$, $r_{comp}$, and $r_{ent}$ as defined in Section~\ref{sec:adv-refine}.
                \STATE Aggregate to $r$ via Equation~\eqref{equ:ra-total}.
                \STATE Update $\boldsymbol{\Theta}_{LLM}$ via GRPO and ~\cite{wang2025beyond} to maximize $r$.
            \ENDFOR
            \IF{LLM backbone converged}
                \STATE Obtain $\boldsymbol{\Theta}_{LLM}^*$
                \STATE \textbf{return} $(\boldsymbol{\Theta}^*_{LLM}, \boldsymbol{\Theta}^*_{R})$
            \ENDIF
        \ENDFOR

    \end{algorithmic}
    
\end{algorithm}

\section{Overall Algorithm of \name}\label{sec:alg}

Here we provide the overall iterative optimization process of \name in Algorithm~\ref{alg:overall}. Specifically, we first freeze the LLM backbone and train Rechead for the downstream recommendation task in Stage I. We then freeze Rechead and use it to provide stable, task-specific feedback for LLM reasoning refinement and alignment in Stage II.

\section{Detailed Experimental Setup}\label{sec:detailed-exp}

\subsection{Dataset}
Following previous works~\cite{you2025r,tan2025reinforced}, we conduct experiments on three datasets from the latest public Amazon source\footnote{https://amazon-reviews-2023.github.io/index.html}, namely Musical Instruments, CDs and Vinyl, and Video Games. Following the temporal-truncation protocol in prior work~\cite{bao2025bi,zhang2025reinforced,you2025r}, we start from interactions in the most recent year to collect at least 10k valid items for each dataset. We omit the 5-core filter to preserve natural recommendation behavior. Each user's interaction history is chronologically ordered and truncated to the latest 20 actions. We then split each dataset into training/validation/test sets with a ratio of 8:1:1 and evaluate on the entire item set. The processed dataset statistics are summarized in Table~\ref{tab:statistics}.

\begin{table}[h]
  \centering
  \caption{Data statistics.}
    \begin{tabular}{cccc}
    \toprule
    Dataset & Users & Items & Interactions \\
    \midrule
    Musical Instruments & 15,656 & 10,320 & 34,373 \\
    CDs and Vinyl & 7,701  & 12,024 & 13,435 \\
    Video Games & 29,230 & 10,144 & 63,502 \\
    \bottomrule
    \end{tabular}%
    \vspace{-3mm}
  \label{tab:statistics}%
\end{table}%

\subsection{Baselines}

Here we briefly introduce the baselines used in Section~\ref{sec:exp}.

\begin{itemize}[leftmargin=*]
    \item \textbf{GRU4Rec}~\cite{hidasi2015session}: RNN-based session recommender with ranking-oriented loss for sequential user behavior modeling.
    \item \textbf{Caser}~\cite{tang2018personalized}: CNN-based model that captures union-level sequential patterns via horizontal and vertical filters.
    \item \textbf{SASRec}~\cite{kang2018self}: Self-attention model for adaptively selecting relevant items from user histories.
    \item \textbf{TIGER}~\cite{rajput2023recommender}: Generative retrieval using semantic IDs and Transformer-based sequence prediction.
    \item \textbf{BIGRec}~\cite{bao2025bi}: LLM-based generative recommender fine-tuned to output tokenized item representations.
    \item \textbf{D$^3$}~\cite{bao2024decoding}: Debiases LLM decoding by controlling ghost tokens and improving output diversity.
    \item \textbf{S-DPO}~\cite{chen2024softmax}: Softmax-enhanced DPO for ranking, leveraging multiple hard negatives.
    \item \textbf{SPRec}~\cite{gao2025sprec}: Self-play framework for DPO training to reduce bias and increase diversity.
    \item \textbf{R$^2$ec}~\cite{you2025r}: Unified reasoning-augmented recommender with dual-head design.
    \item \textbf{ReRe}~\cite{tan2025reinforced}: RL-based generative recommender with verifiable rewards and constrained beam search.
    \item \textbf{LatentR$^3$}~\cite{zhang2025reinforced}: RL-optimized latent reasoning without explicit chains for efficient LLM-based recommendation.
\end{itemize}

\subsection{Implementation Details}\label{sec:imp}
We train \name on 4 GPUs. To ensure fair comparison, we set the batch size to 256 for non-LLM methods and 8 for LLM-based methods. All LLM-based methods, including \name, use Qwen3-0.6B~\cite{qwen3technicalreport} as the backbone model for efficient generation. The maximum generation length of all LLMs is set to 768 for efficiency. We use static-retrieval-mrl-en-v1~\cite{reimers-2019-sentence-bert,kusupati2024matryoshka,henderson2017efficient} as the sentence transformer encoder within the Rechead module. For each model, the model-specific hyperparameters are tuned via grid search, while shared hyperparameters remain fixed across experiments. We report the average performance over 10 runs using the optimal hyperparameter configuration.

\section{Limitations}

Although \name is effective, there remains room to further improve the quality of the generated reasoning CoTs. For example, introducing additional optimization signals, such as diversity-oriented metrics and hallucination mitigation strategies, may further enhance reasoning quality and potentially yield additional recommendation gains. We consider these directions promising avenues for future work.

%% file: 2preliminary.tex
\section{Preliminary}
\label{sec:preliminary}

This section introduces the two foundations of our study: Group Relative Preference Optimization (GRPO)~\cite{shao2024deepseekmath}, a representative method in Reinforcement Learning with Verifiable Rewards (RLVR), and general LLM-based recommendation task formulation, which reformulates recommendations as a language modeling task.

\subsection{Group Relative Preference Optimization (GRPO)}

Group Relative Preference Optimization (GRPO)~\cite{shao2024deepseekmath} is an efficient algorithm for aligning LLMs under verifiable reward supervision. For a policy $\pi_{\theta}$ that generates textual output $o$ given an input $x$, GRPO samples $G$ candidate responses $\{o_i\}_{i=1}^{G}$ and obtains a scalar reward $r_i$ for each. The reward advantage is normalized within the group $G$:
\begin{equation}
\hat{A}_i = \frac{r_i - \mu_G}{\sigma_G},
\end{equation}
where $\mu_G$ and $\sigma_G$ are the mean and standard deviation of $G$ rewards. At the token level, the ratio between the trainable policy $\pi_{\theta}$ and a fixed reference model $\pi_{\mathrm{ref}}$ is computed as
\begin{equation}
w_{i,t}(\theta) = \frac{\pi_{\theta}(o_{i,t} \mid x, o_{i,<t})}
{\pi_{\mathrm{ref}}(o_{i,t} \mid x, o_{i,<t})},
\end{equation}
where $o_{i,t}$ is the $t$-th token of $o_i$. The optimization objective is then
\begin{equation}
\begin{aligned}
\mathcal{J}_{\mathrm{GRPO}}(\theta) = &
\mathbb{E}[
\frac{1}{G}\sum_{i=1}^{G}\frac{1}{|o_i|}\sum_{t=1}^{|o_i|}
\min\!(w_{i,t}(\theta)\hat{A}_i,\\
&\mathrm{clip}(w_{i,t}(\theta),1-\varepsilon,1+\varepsilon)\hat{A}_i)
- \beta\,\mathbb{D}_{\mathrm{KL}}\!\left[\pi_{\theta}\|\pi_{\mathrm{ref}}\right]
],
\end{aligned}
\end{equation}
where $\mathbb{D}_{\mathrm{KL}}\!\left[\pi_{\theta}\|\pi_{\mathrm{ref}}\right]$ is the estimated KL divergence between $\pi_{\theta}$ and $\pi_{\mathrm{ref}}$, and $\varepsilon$ is the clipping ratio. GRPO thus adjusts token probabilities using relative in-group feedback, providing trajectory sampling-based preference alignment, achieving stable policy improvement without explicit value estimation.

Recent work extends GRPO from an averaged token-level perspective by introducing entropy-guided optimization~\cite{wang2025beyond}. This approach observes that only a minority of tokens that exhibit high uncertainty are critical “decision tokens” influencing reasoning outcomes. The token entropy is therefore defined as
\begin{equation}
E_t = -\sum_{\hat{o}_t \in vocab} \pi_{\theta}(\hat{o}_t \mid x, o_{<t}) \log \pi_{\theta}(\hat{o}_t \mid x, o_{<t}),
\end{equation}
which measures uncertainty in token selection at token position $t$ with token probabilities on the whole vocabulary ($vocab$). To enhance policy learning efficiency, the reward updates are restricted to the subset of tokens with the highest entropy. Let $\mathcal{S} = \{t \mid E_t \geq \tau_\rho^{\mathcal{B}}\}$ denote the indices of tokens exceeding the entropy threshold $\tau_\rho^{\mathcal{B}}$, which is the top $\rho$ threshold within (mini-)batch $\mathcal{B}$. The updates are thus computed only for the tokens in $\mathcal{S}$ rather than across all token positions. This selective gradient approximation effectively focuses learning on the most informative portions of reasoning trajectories, reducing computational cost and noise from low-entropy tokens. Empirically, such selective reinforcement—optimized with top $\rho=20\%$ of tokens--achieves comparable alignment performance while improving stability and interpretability. The entropy-guided refinement thus deepens theoretical understanding of GRPO and suggests scalable strategies for reinforcement learning in large language models.

\subsection{LLM-based Recommendations}

LLM-based recommendation~\cite{zhao2024let, acharya2023llm} leverages LLMs to perform recommendations through natural-language understanding and generation. Given a user’s interaction history $H_u = [v_1, \dots, v_J]$, each item $v_j$ is described textually, and the history is composed into a prompt $x_u$ and is encoded with the LLM's last hidden state into $\boldsymbol{h}_u$. For a candidate item $v$, similar representation $\boldsymbol{h}_v$ is obtained from encoding item descriptions $x_v$, and their matching score can be computed as
\begin{equation}
s(u, v) = \boldsymbol{h}_u^\top \boldsymbol{h}_v.
\end{equation}

Contrastive learning is then applied to separate positive from negative items during training~\cite{gao2025sprec,you2025r}.

\begin{equation}
\mathcal{L}_{\mathrm{CL}}=-\log \frac{\exp \left(s(u, v^+) / \tau\right)}{\sum_{v^{\prime} \in B_{v^+}} \exp \left(s(u, v^{\prime}) / \tau\right)},
\end{equation}
where $\tau$ is a temperature parameter, $B_{u, v^+}$ denotes the union of positive and negative samples for the interaction $(u, v^+)$, and $v^+$ is the ground-truth positive item. When $\tau=1$, this formulation reduces to the standard Cross-Entropy (CE) loss. However, directly optimizing LLM hidden states with task-specific objectives may disrupt the model's inherent reasoning structures, resulting in degraded reasoning capability and suboptimal recommendation performance.

Another paradigm, generative variants~\cite{tan2025reinforced}, instead let the model autoregressively produce the next preferred item, modeling
\begin{equation}
P(o \mid x_u) = \prod_{t=1}^{T} P(o_t \mid x_u, o_{<t}),
\end{equation}
where the generated text represents the reasoning trace, predicted item text, or item SIDs. This formulation unifies recommendation with natural-language generation, enabling reasoning-aware and interpretable preference modeling based on the power of LLMs. However, a mismatch between natural language expressions and discrete item identifiers forces a subsequent approximate search or constraint decoding that compromises accuracy. Even with SIDs for generative recommendation, semantic gaps persist due to differences in tokenization and training between the LLM’s native vocabulary and the SID space used in recommendation.